\documentclass[aps,pre,onecolumn,showpacs]{revtex4}
\usepackage{graphicx}
\usepackage{wasysym}

\newcommand{\be}{\begin{equation}}
\newcommand{\ee}{\end{equation}}
\newcommand{\bea}{\begin{eqnarray}}
\newcommand{\eea}{\end{eqnarray}} 
\newcommand{\km}{\langle k \rangle} 

\begin{document}

\title{Comparison of voter and Glauber ordering dynamics on networks}

\author{Claudio Castellano}
\email{castella@pil.phys.uniroma1.it}
\affiliation{Dipartimento di Fisica, Universit\`a di
Roma ``La Sapienza'' and Center for Statical Mechanics and Complexity,
INFM Unit\`a Roma 1, P.le A. Moro 2, I-00185 Roma, Italy}

\author{Vittorio Loreto}
\email{loreto@roma1.infn.it}
\affiliation{Dipartimento di Fisica, Universit\`a di
Roma ``La Sapienza'' and Center for Statical Mechanics and Complexity,
INFM Unit\`a Roma 1, P.le A. Moro 2, I-00185 Roma, Italy}

\author{Alain Barrat}
\email{Alain.Barrat@th.u-psud.fr}
\affiliation{Laboratoire de Physique Th\'eorique (UMR 8627 du CNRS),
B\^at. 210, Universit\'e de Paris-Sud, 91405 Orsay Cedex, France}

\author{Federico Cecconi}
\email{f.cecconi@istc.cnr.it}
\affiliation{Institute of Cognitive Sciences and Technologies C.N.R.,
Viale Marx, 15, 00137, Roma, Italy}

\author{Domenico Parisi}
\email{d.parisi@istc.cnr.it}
\affiliation{Institute of Cognitive Sciences and Technologies C.N.R.,
Viale Marx, 15, 00137, Roma, Italy}

\date{\today}


\begin{abstract}
We study numerically the ordering process of two very simple dynamical models
for a two-state variable on several topologies with increasing
levels of heterogeneity in the degree distribution.
We find that the zero-temperature Glauber dynamics for the Ising model
may get trapped in sets of partially ordered metastable states even for finite
system size, and this becomes more probable as the size increases.
Voter dynamics instead always converges to full order on finite networks,
even if this does not occur via coherent growth of domains.
The time needed for order to be reached diverges with the system size.
In both cases the ordering process is rather insensitive to the variation
of the degreee distribution from sharply peaked to scale-free.
\end{abstract}

\maketitle
\section{Introduction}
Complex networks have become astonishingly popular in recent years
as models for the interaction patterns among individuals
(agents/computers/molecules/etc...) in many diverse fields~\cite{Newman}.
The nontrivial topologies found in nature are often the structures
over which dynamical processes take place.
Complex interaction patterns produce in many cases
nontrivial collective dynamical properties, very different from those
observed when single elements are connected in a regular way.

Social systems are one of the fields where it is more evident
that regular lattices are often inappropriate as models
for the structure of interactions.
This has led to intense activity aimed at investigating the effect
of complex networks on the behavior of models for the spreading of
opinions~\cite{opinions}, the diffusion of culture~\cite{Axelrod97}
and other processes where domains of homogeneous individuals emerge
out of an initial disordered state~\cite{Kuperman02,Klemm03,Stauffer04}.
These studies have often been motivated by a direct interest in the social
phenomena described by the models considered.
Here we take a different point of view. We try to understand how
the basic features of ordering processes occurring on complex networks
depend on the topology and whether generic rules
can be inferred.

For this reason we consider the simplest models that exhibit
an ordering dynamics: the voter model and the Glauber-Metropolis
zero-temperature dynamics for the Ising model.
They are not intended to describe any real phenomenon, rather they
allow the investigation of the role of simple physical ingredients
(as surface tension or interfacial noise), and their interplay with
the topological structure, in determining the overall behavior of the system.

For the same reason we consider extremely simple networks:
the complete graph, the random graph and the scale-free Barabasi-Albert graph.
We believe that our investigation provides the necessary background
for a thorough understanding of more complicated and realistic dynamics
on more complex networks.

As a general pattern, we find that the two types of dynamics behave rather
differently on networks, while they are not much sensitive
to the precise topology on which they evolve.
For systems with a finite number of sites $N$, Glauber zero-temperature
dynamics leads in some realizations to full ordering, while in others
the system gets trapped in a set of disordered metastable configurations.
The probability of not reaching order tends to grow with $N$.
The voter dynamics instead always reaches the fully ordered state
when the system size is finite. This however does not involve a
coherent ordering process as it happens on regular lattices: the system
remains on average disordered until a random fluctuation leads it to
consensus.
The temporal scale $\tau(N)$ needed for this process diverges as $N^\gamma$.
The detailed form of the topology affects only the value of $\gamma$.

The paper is organized as follows. In the next Section we describe the
models studied and we give a brief overview of the background.
Section~\ref{voter} reports the results for voter dynamics, on the
various networks considered.
Glauber zero-temperature dynamics is discussed in Section~\ref{Glauber}.
The final Section contains a summary and a discussion of the results.

\section{The models}

In this paper we consider two of the simplest models that, starting
from a fully disordered initial state, exhibit an ordering dynamics.
In both cases a single variable $s_i$, that may assume only two values
($\pm 1$), fully specifies the state (opinion, culture) of site $i$.

In the zero-temperature Glauber dynamics for the Ising model,
at each time step a node $i$ selected randomly and the local field is computed.
Then $s_i$ is set equal to +1 if the local field is positive and to $-1$
if it is negative. If the field is zero the variable changes its
value~\cite{note1}.

While the Glauber dynamics at finite temperature reaches
asymptotically the state dictated by thermodynamic equilibrium, which is
well known also for some complex networks~\cite{Leone02}, the
situation at zero temperature is much less understood.
On regular lattices it is known that, while in $d=1$
a fully ordered state is always reached, in higher dimensions the
system may get stuck in a frozen state with coexisting domains of
opposite magnetization.
In $d=2$ this occurs with probability of about 1/3 for large systems.
In higher dimensions the probability of reaching the ground state rapidly
vanishes as the system size grows and the system ends up wandering forever
within an iso-energy set of metastable states~\cite{Spirin01}.
The Glauber zero-temperature dynamics has been recently studied by
Boyer and Miramontes~\cite{Boyer03} on the Watts-Strogatz small-world
network~\cite{Watts98}.
They observe that ordering is hindered by the presence of shortcuts
leading to a pinned state with a finite size of ordered domains.

The other type of dynamics we consider is the voter model:
at each time step a node ($i$) and one of its neighbors ($j$) are randomly
selected. Then $s_i$ is set equal to $s_j$.
Also in this case the behavior on regular lattices is well
known~\cite{Krapivsky92,Rednerbook}: in $d \le 2$ order is reached
asymptotically and the density of interfaces decays as $t^{(d-2)/2}$;
for $d>2$ a stationary active state is reached, with coexisting domains.
Also this model has been studied on the small-world network of Watts and
Strogatz finding that the dynamics gets stuck in 
a disordered stationary state on infinitely-large systems, while it orders
in finite systems over a time scaling as $N$~\cite{Castellano03}.
Recently, Wu and Huberman~\cite{Wu04} and
Suchecki {\em et al.}~\cite{Suchecki04} have considered the voter dynamics
on networks with heterogeneous node degrees, showing that the average
magnetization is not conserved, at odds with what occurs on regular
lattices.
During the completion of this work a preprint by Sood and Redner~\cite{Sood04}
has appeared, reporting an analytical investigation of the ordering of
the voter model on heterogeneous networks. They compute the time to
reach full consensus as a function of $N$ for networks with generic
degree distribution.

Glauber dynamics leads to the formation of ordered domains and their
coarsening as a consequence of the existence of surface tension and the
drive provided by energy minimization.
In the voter dynamics instead no surface tension exists and the tendency
toward order is the effect of annihilation of freely diffusing interfaces
between domains~\cite{Dornic01}.

In the present paper the dynamical behavior of these models is studied
on different types of topology.
For random graph we intend an Erdos-Renyi graph: a set of $N$ nodes
such that between each pair of them there is a connection with probability
$p$~\cite{Bollobasbook}.
We have always checked that all nodes considered formed a connected cluster.
For relatively small average degree $\km>1$ and large system
size $N$, while a giant component exists, including the overwhelming
majority of nodes, isolated smaller clusters may appear.
We have discarded them and run the dynamics only on the giant component
of the network, always checking that at least 98\% of the $N$ nodes
belonged to it and that the average degree was very close to its nominal
value $\km=p(N-1)$.
To study the effect of the scale-free topology we consider the most
common model known to produce such type of structure: the Barabasi-Albert
graph~\cite{Barabasi99}.

To monitor the ordering dynamics, we focus on the
density of active bonds $n_A(t)$, i.e. bonds connecting sites with
opposite values of the variable $s_i$. For the Ising model it coincides
with the energy density.
The initial condition is always given by a completely disordered system,
i.e. variables are chosen to be $-1$ or $+1$ at random in a completely
uncorrelated way.
Results presented are always averaged over a large number of realizations
of both the topology and the initial condition.

\section{Voter model}
\label{voter}

\subsection{Complete graph}
We start by considering a random graph in the limit $p=1$, i.e. a complete
graph.
In such a limit the graph is fully connected and the state of the whole system
is fully specified by only one variable, the magnetization
$\mu = \langle s_i \rangle$.

On such a graph, as well as on a regular lattice, the voter model
coincides with the so called ``Ochrombel simplification of the Sznajd model'',
for which exact results have been derived by Slanina and
Lavicka~\cite{Slanina03}.
Writing down the Master Equation for the probability density $P(\mu,N_a)$
of having magnetization $\mu$ after $N_a$ attempted updates and passing
to the continuum limit one obtains
\be
{\partial P(\mu,t') \over \partial t'} = {\partial^2 \over \partial \mu^2}
[(1-\mu^2) P(\mu,t')],
\label{eqvoter}
\ee
where the natural scaling of time with the number of sites $N$,
$t' = N_a/N^2$, has been introduced.

Eq.~(\ref{eqvoter}) is a one-dimensional diffusion equation with a
variable diffusion coefficient.
It is solved by standard methods for the Fokker-Planck
equation~\cite{Slanina03}, finding that, for large $t'$, the fraction
of bonds connecting nodes with opposite values of the variable
(active bonds) is
\be
n_A(t') = {(1-m_0^2)\over 2} e^{-2t'},
\label{navoter}
\ee
where $m_0$ is the initial magnetization.

In the following we will measure time
as the number of attempted updates per node $t = N_a/N$, according to
the idea that each individual tries to modify its state once per unit time.
In this way the voter dynamics on a complete graph has a characteristic
time $\tau(N)=N/2$.

\begin{figure}
\includegraphics[angle=0,width=10cm,clip]{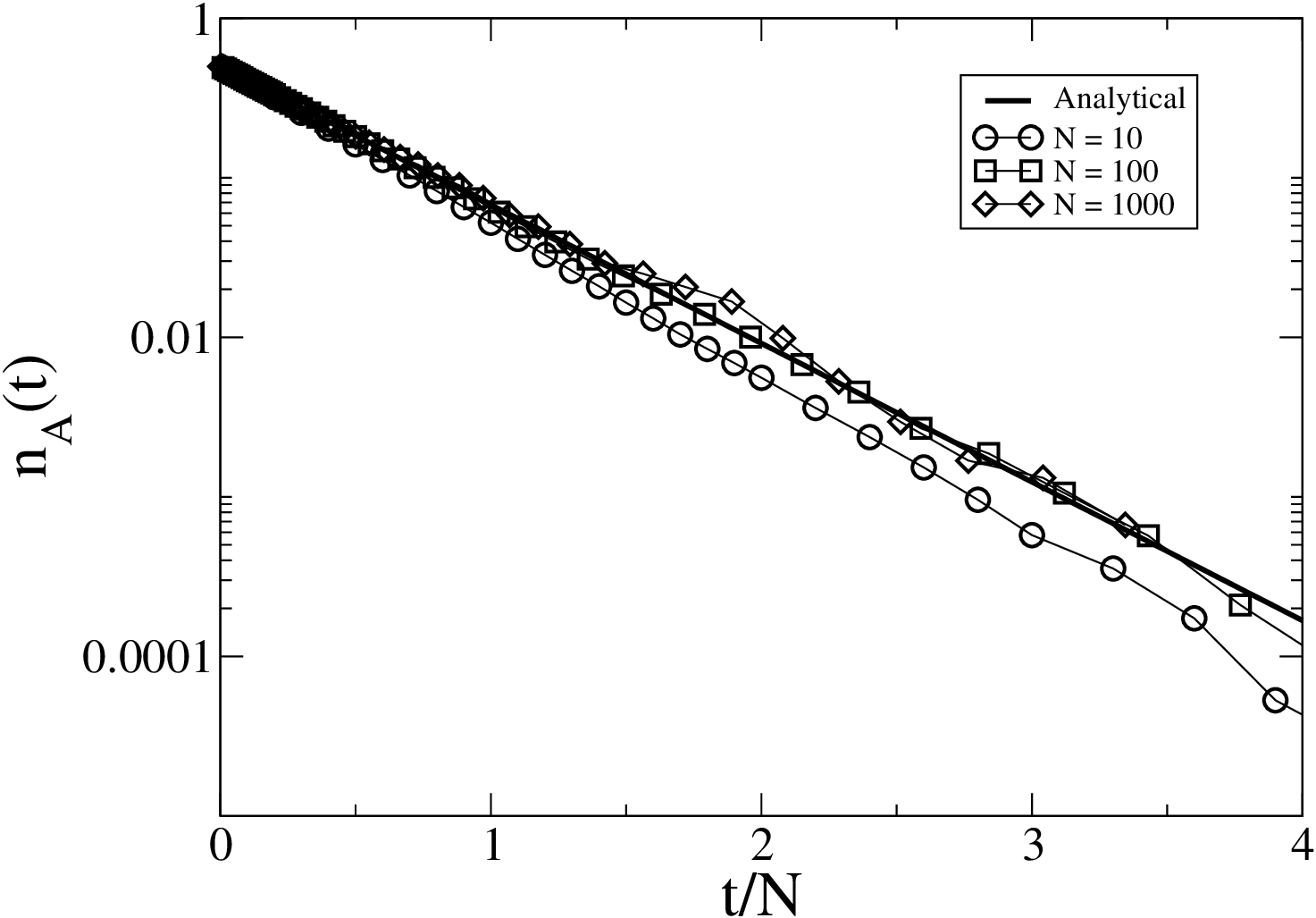}
\caption{Fraction of active bonds for voter dynamics on a complete graph.}
\label{Fig6}
\end{figure}

These results are exacts only in the limit $N \to \infty$.
As discussed in Ref.~\cite{Slanina03}, for finite $N$
diffusive terms proportional to $1/N$ appear in the expansion of the
Master Equation.
Nevertheless, numerical simulations (Fig.~\ref{Fig6}) show that
Eq.~(\ref{navoter}) perfectly describes the evolution of the system
starting from relatively small values of $N$.
 
In order to gain further insight, let us separately consider the survival
probability $\rho(t)$, i.e.
the probability that a run has not reached the fully ordered state up to
time $t$ and $n_A^S(t)$, the fraction of active bonds {\em averaged only over
surviving runs}. Clearly the equality $n_A(t) = \rho(t) n_A^S(t)$ holds.

The survival probability is evaluated in Ref.~\cite{Slanina03} and it reads,
for $m_0=0$
\be
\rho(t) = \left\{
\begin{array}{cll}
1                 & & t \ll \tau(N) \\
{3 \over 2} e^{-t/\tau(N)}  & & t \gg \tau(N) .
\end{array} \right.  \:
\label{eqrho}
\ee
The quantity $n_A^S(t)$ is easily computed and turns out to be
\be
n_A^S(t) = \left\{
\begin{array}{cll}
{1 \over 2} e^{-t/\tau(N)}   & & t \ll \tau(N) \\
{1 \over 3}                  & & t \gg \tau(N) .
\end{array} \right.  \:
\label{eqnast}
\ee

We realize then that the fully ordered state is not reached in the
thermodynamic limit. This occurs for two reasons. The first is that
the temporal scale $\tau(N)$ over which consensus is reached in finite
systems {\em diverges} with the size $N$. This happens also on
regular lattices and is already evident from the behavior of $n_A(t)$.
The second reason, specific to graphs, is that even for $t \gg \tau(N)$
the fraction of active bonds in surviving runs $n_A^S(t)$ does not go to zero
when $N$ grows.
This means that surviving runs do not order; they rather stay
in configurations with, on average, a finite (and large) fraction
of active bonds. Random fluctuations bring eventually all surviving runs
to the fully ordered absorbing state; however, as long as the runs survive
they do not order on average.
The decay of $n_A(t)$ is just a consequence of the decay of $\rho(t)$,
the number of such surviving runs.

This is completely different from what occurs on regular lattices.
In such a case $\rho(t)$ remains 1 up to a long time, after which
it quickly goes to zero. The decay of $n_A(t)$ mirrors
the decay of $n_A^S(t)$: all runs survive for approximately the same time
and they all get more and more ordered.
An example of such fully ordering behavior is provided by Glauber dynamics
on the complete graph (see below, Fig.~\ref{Figx}).

\subsection{Random graph}
\begin{figure}
\includegraphics[angle=0,width=10cm,clip]{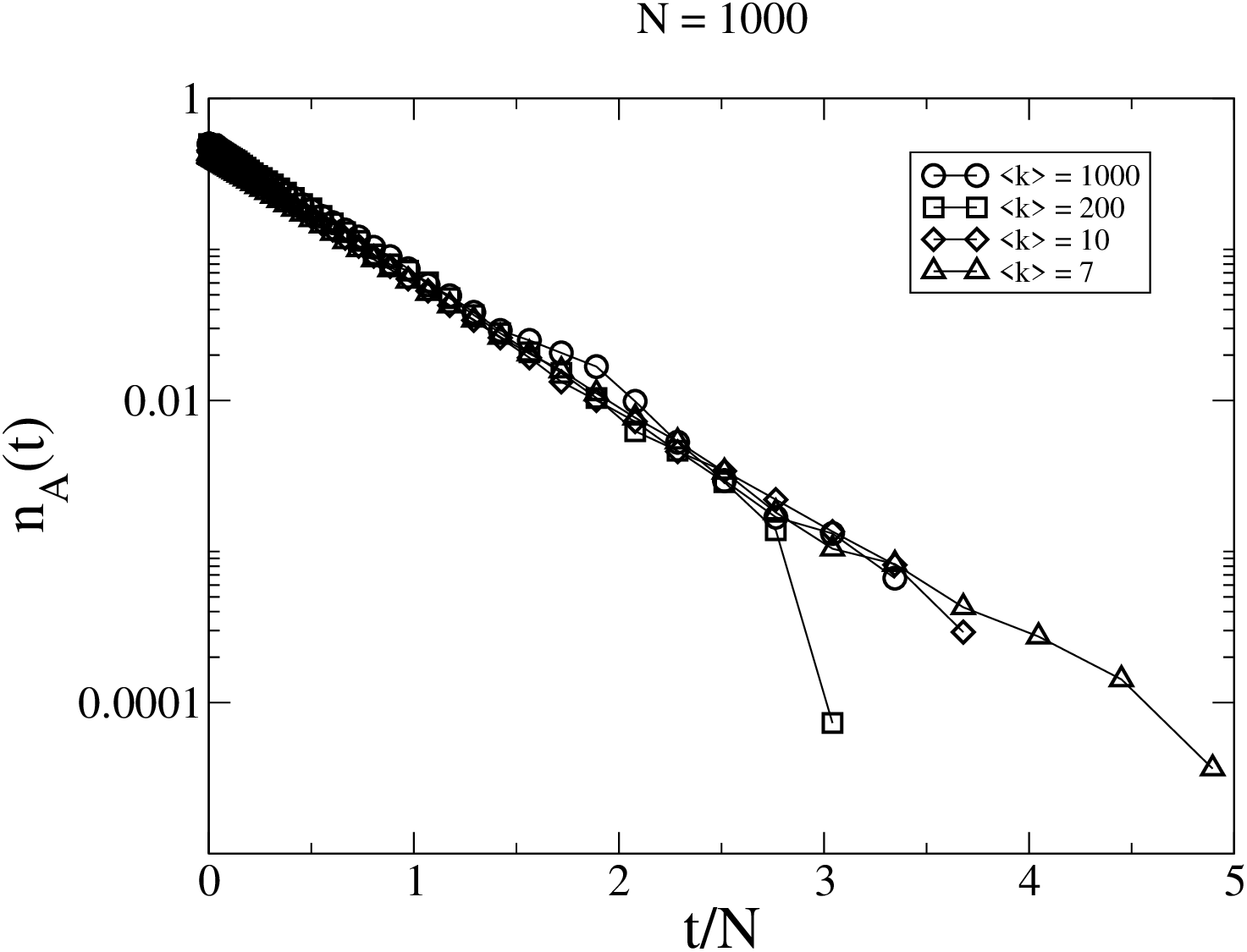}
\caption{Fraction of active bonds for voter dynamics on a random graph.}
\label{Fig7}
\end{figure}

Let us now consider what occurs for fixed $N$ and changing $p$, that
is, the average degree $\km=p(N-1) \approx pN$ of nodes
(Fig.~{\ref{Fig7}}).
We find a remarkable similarity of the temporal evolution with the
case of the complete graph, even when the average degree is changed
by a factor larger than 100. Only the prefactor weakly depends on $\km$.
The characteristic time scale $\tau(N)$ is proportional to $N$, as
found in Ref.~\cite{Sood04},
and independent from the average degree $\km$.

\begin{figure}
\includegraphics[angle=0,width=10cm,clip]{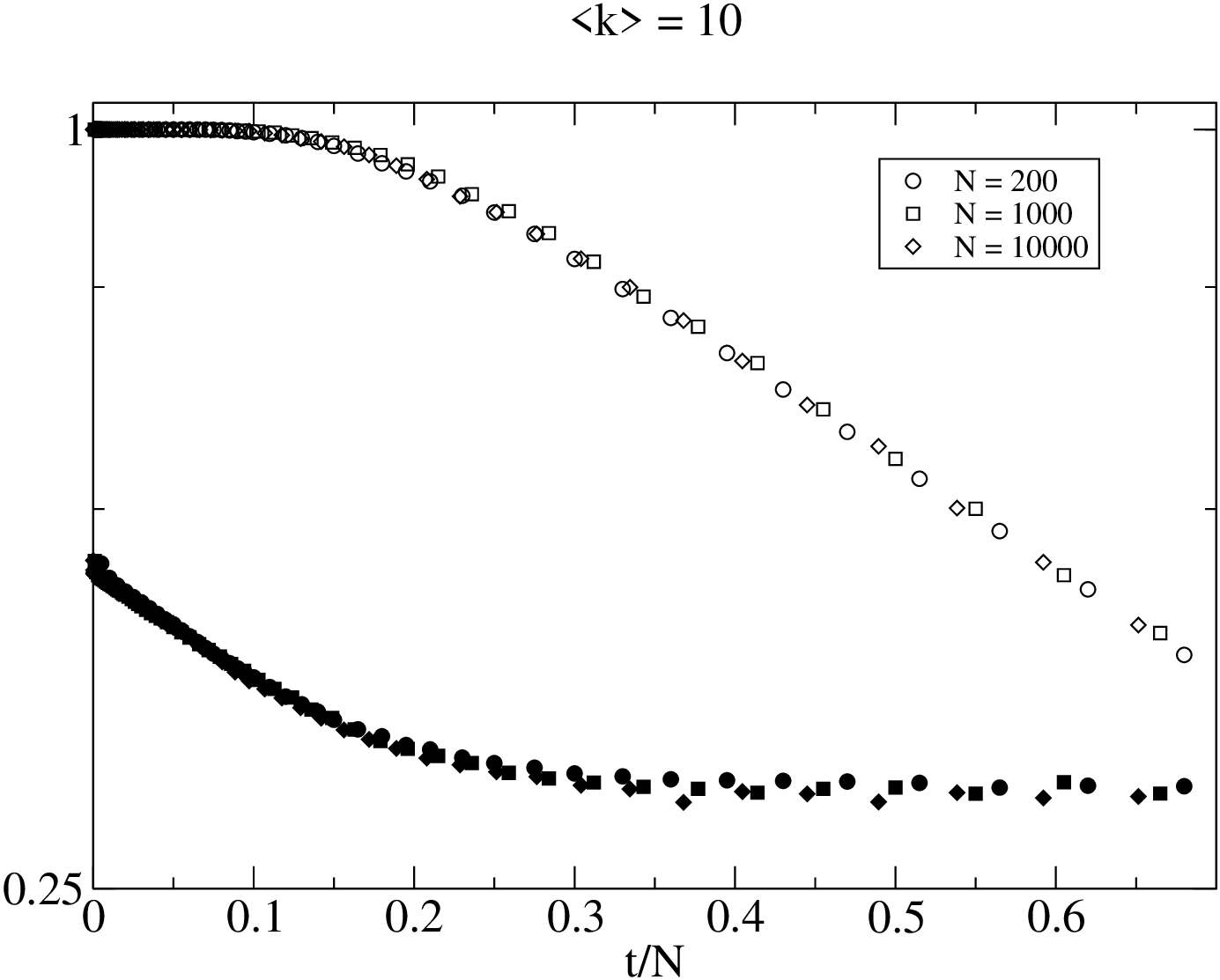}
\caption{Fraction of active bonds in surviving runs $n_A^S(t)$
(filled symbols) and survival probability $\rho(t)$ (empty symbols)
for voter dynamics on a random graph for $\km=10$.}
\label{Fig7bis}
\end{figure}

Figure~{\ref{Fig7bis}} reports the value of $n_A^S$ and $\rho(t)$ as a
function of $t/N$ for several values of $N$ and $p$, with constant
average degree $\km=10$.
The behavior is very similar to what happens on a complete graph
and can be summarized as follows
\be
\rho(t) = \left\{
\begin{array}{cll}
1                 & & t \ll \tau(N) \\
{3 \over 2} e^{-t/\tau(N)}  & & t \gg \tau(N),
\end{array} \right.  \:
\ee
\be
n_A^S(t) = \left\{
\begin{array}{cll}
{3 \over 2} A(\km) e^{-t/\tau(N)}   & & t \ll \tau(N) \\
A(\km)                  & & t \gg \tau(N) .
\end{array} \right.  \:
\ee
so that $\rho(t) = 3/2 A(\km) e^{-t/\tau(N)}$ for all times.
The prefactor $A(\km)$ is equal to $1/3$ for the fully connected graph,
while it is smaller for finite $\km$.
We can conclude that, on random as on complete graphs, surviving runs
do not get ordered.

\subsection{Barabasi-Albert graph}

We consider then the voter model evolving on a scale-free graph built
according to the rules of Barabasi and Albert (BA)~\cite{Barabasi99}.
These graphs are constructed by considering an initial fully connected 
core of $m+1$ nodes and iteratively adding new nodes, each with $m$ bonds.
The other node to which a new bond is linked is chosen among existing
nodes with probability proportional to their degree
(preferential attachment).

\begin{figure}
\includegraphics[angle=0,width=10cm,clip]{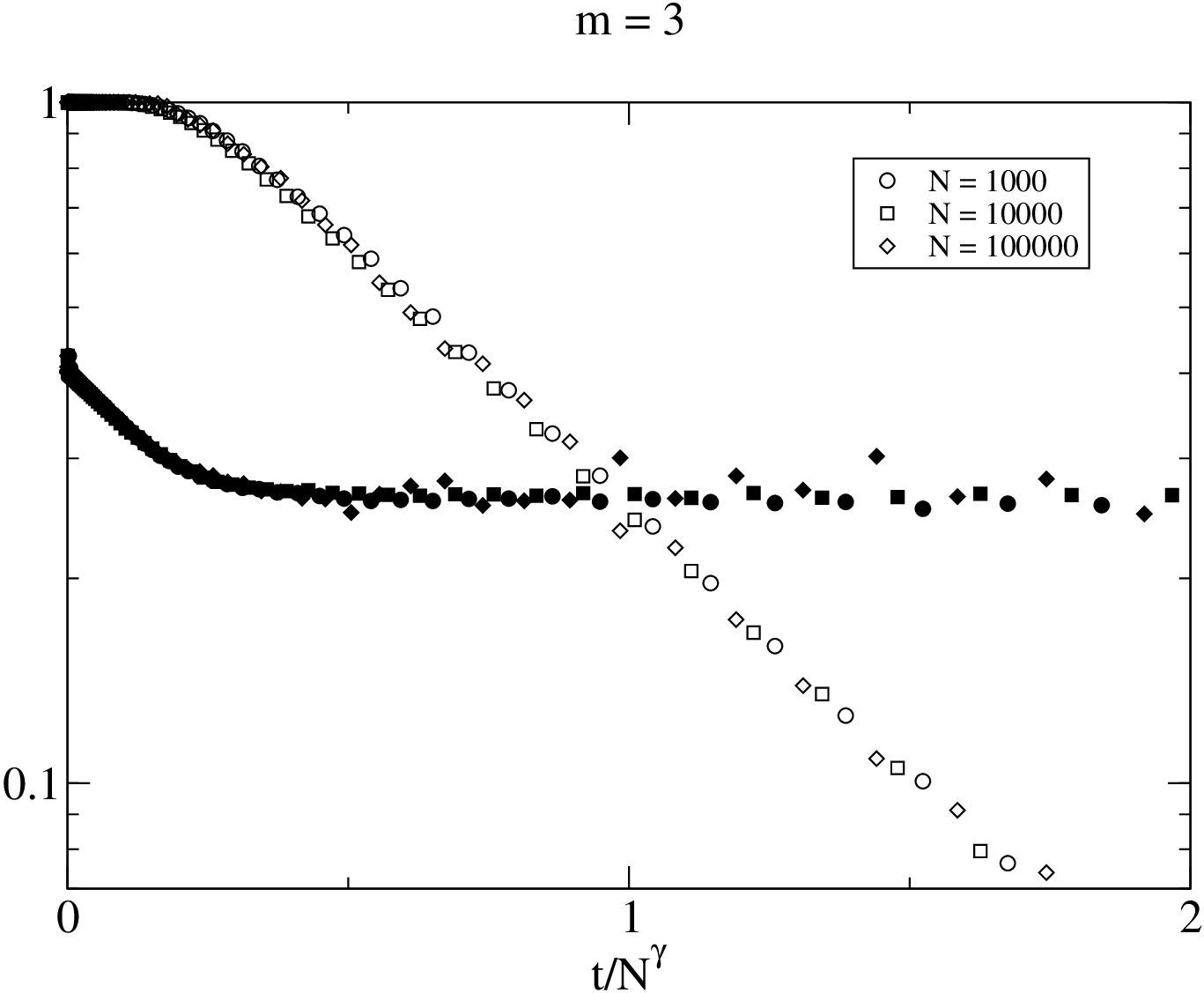}
\caption{Fraction of active bonds in surviving runs $n_A^S(t)$
(filled symbols) and survival probability $\rho(t)$ (empty symbols)
for voter dynamics on a Barabasi-Albert graph for $m=3$.}
\label{Figx4}
\end{figure}

\begin{figure}
\includegraphics[angle=0,width=10cm,clip]{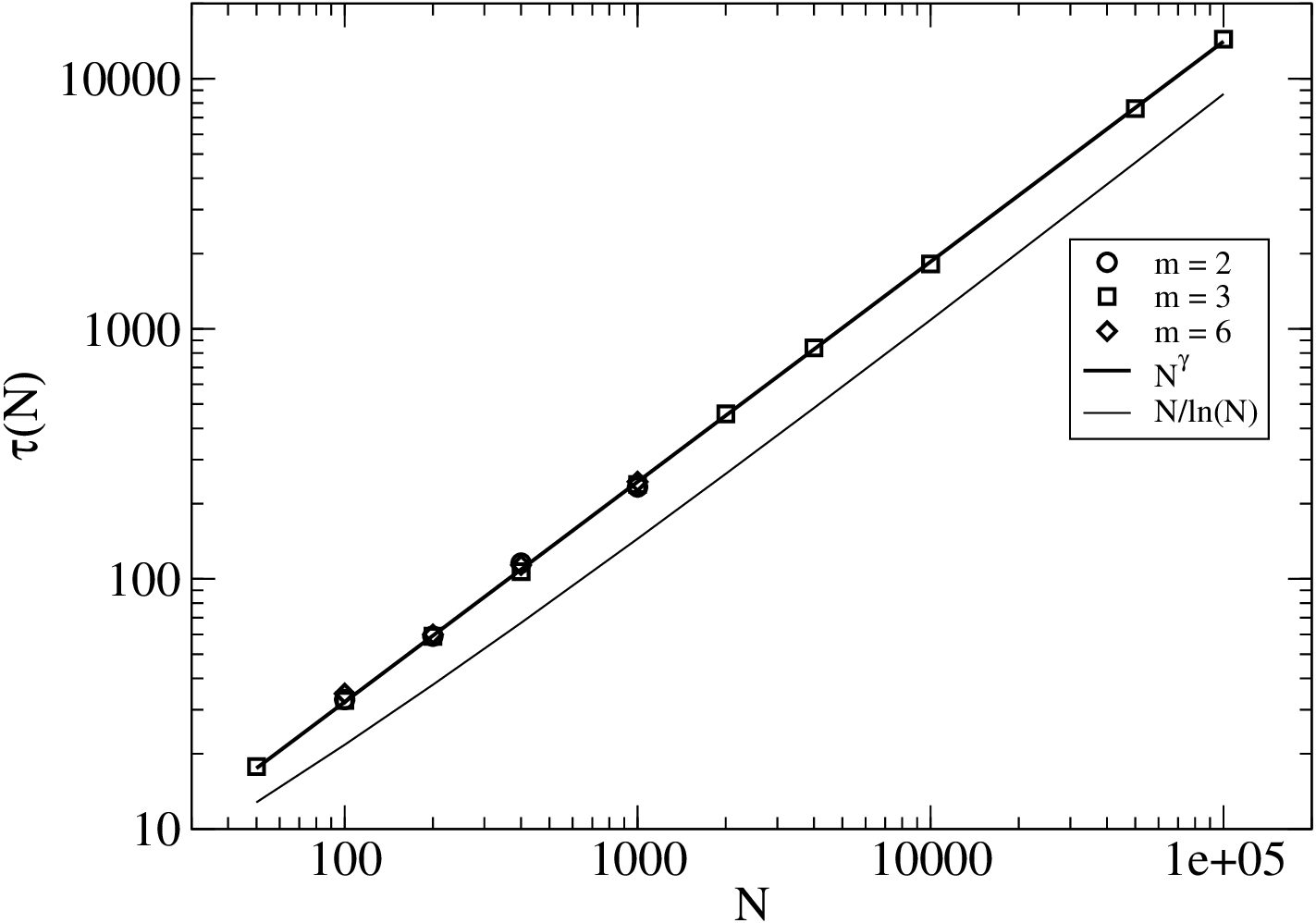}
\caption{Scaling of the time $\tau(N)$ vs $N$ on Barabasi-Albert graphs
for several values of $m$, with a pure power-law fit with $\gamma=0.880$
(thick line) and the formula $N/\log(N)$ (thin line).}
\label{Fig8}
\end{figure}

Also on the BA networks the fraction of active bonds
$n_A(t)$ as a function of time decays, for $m>1$, exponentially fast to zero.
The survival probability $\rho(t)$ and the
fraction of active bonds restricted to surviving runs $n_A^S(t)$
(Fig.~\ref{Figx4}) follow Eqs.~(\ref{eqrho}) and~(\ref{eqnast}),
i.e. on finite systems the model always reaches the perfectly ordered state,
but surviving runs do not order.
The only difference is the scaling of time with the number $N$ of
nodes, which is reported in Figure~\ref{Fig8}.
A power-law fit yields, independently from $m>1$,
$\tau(N) \sim N^\gamma$, with $\gamma = 0.880 \pm 0.003$.
The nontrivial scaling of $\tau(N)$ with $N$ had already been observed
by Suchecki {\em et. al.}~\cite{Suchecki04}, which estimated $\gamma$
by fitting $\tau(N)$ over a decade. Here we find a compatible value
over more than 3 decades.
In Ref.~\cite{Sood04}, Sood and Redner estimate analytically
$\tau(N)=N/\log(N)$ for this case. In Figure~\ref{Fig8}, we compare
this expression to numerical data, finding a good agreement,
but no sign of the increase of the effective exponent, which would be
the signature of the logarithmic correction.

\begin{figure}
\includegraphics[angle=0,width=10cm,clip]{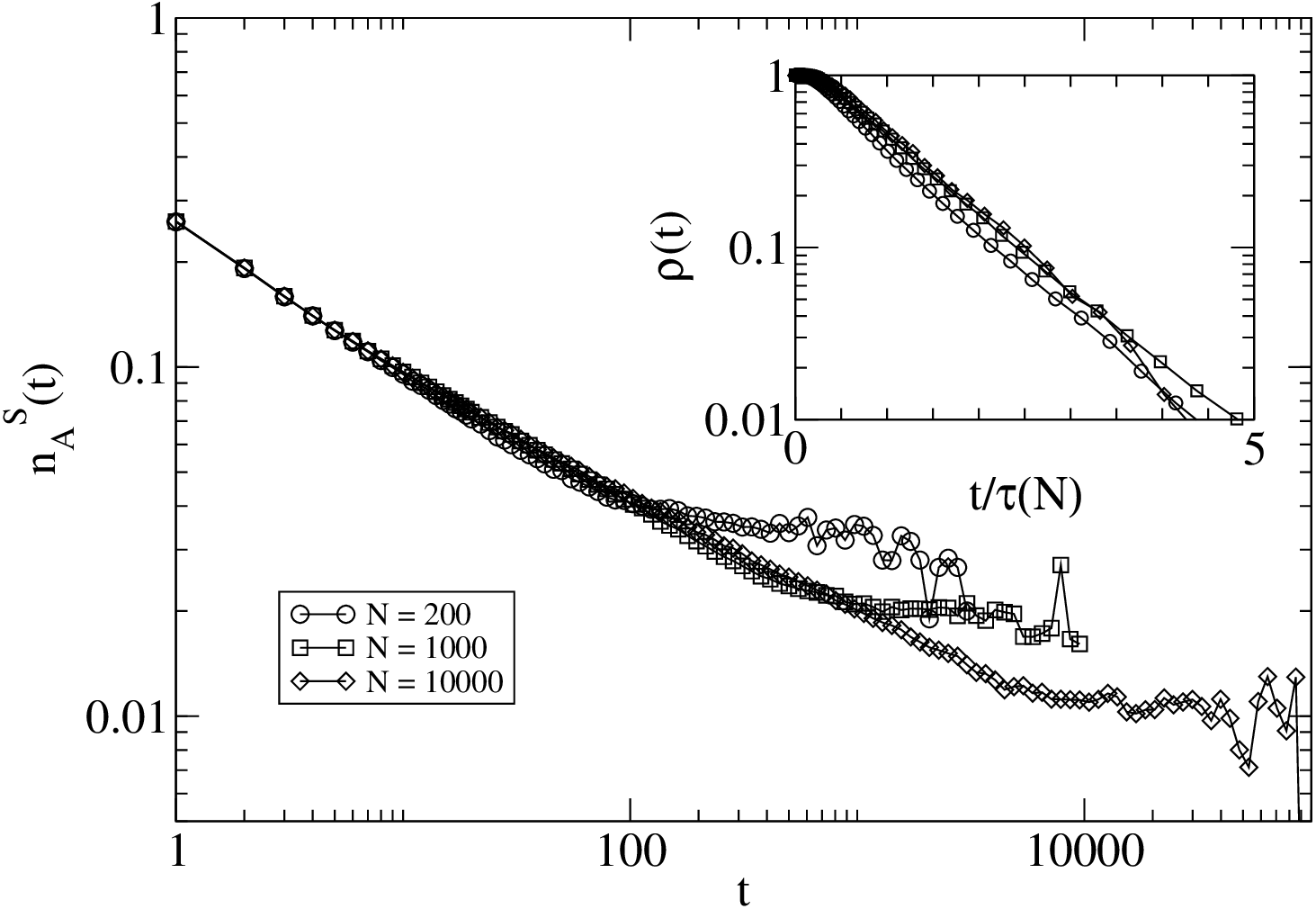}
\caption{Fraction of active bonds in surviving runs $n_A^S(t)$
(main) and survival probability $\rho(t)$ (inset)
for voter dynamics on a Barabasi-Albert graph with $m=1$.}
\label{Fig8bis}
\end{figure}

Also for $m=1$ consensus is reached on finite systems, but much more
slowly: the decay of $n_A^S(t)$ is not exponential; it is power-law or,
possibly, even slower since the exponent is close to $1/3$
on the accessible temporal scales but it seems to be decreasing with time
(Fig.~\ref{Fig8bis}).
Furthermore the plateau of $n_A^S(t)$ weakly depends on $N$.
This is probably a preasymptotic effect.
The survival probability $\rho(t)$ (Fig.~\ref{Fig8bis}, inset)
decays exponentially over a temporal scale $\tau(N) \sim N^\gamma$
with $\gamma=1.04 \pm 0.01$.
The differences with respect to the case $m>1$ are a consequence
of the tree-like structure of the BA network for $m=1$.

In order to investigate the universality of the exponent $\gamma$
we finally study the ordering dynamics of the voter model
on a network built according to the prescriptions of
Ref.~\cite{Dorogovtsev01}.
This graph is grown by iteratively adding nodes. Each of them
is connected to the nodes linked by $m$ randomly chosen
edges. In this way a preferential attachment mechanism is implemented
so that this network has topological properties
practically identical to the one by Barabasi-Albert, with the notable
exception of a large clustering coefficient~\cite{Barrat04}.
This variation has little impact on the ordering dynamics.
For $m>1$, the phenomenology is exactly the same of the BA graph, with the
sole difference that $\gamma=0.978 \pm 0.005$.

In summary we find that the voter dynamics on the scale-free networks
with $m>1$ shows a remarkable similarity with the dynamics on random
and complete graphs.
The nontrivial topology of the BA graph is reflected only in the
scaling of the characteristic time with the number of sites,
which follows a different power-law.

\section{Glauber $T=0$ dynamics}
\label{Glauber}

\subsection{Complete graph}
On the complete graph, also Glauber dynamics can be solved analytically in the
limit $N \to \infty$.
The Master Equation for the probability $P(q,t)$ of having a
fraction $q$ of positive spins at time $t$, is, in the continuum
limit, for $q>1/2$,
\be
{\partial P(q,t) \over \partial t} = - {\partial \over \partial q}
[(1-q) P(q,t)],
\ee
where the natural definition of time is $t = N_a/N$.

The ansatz $P(q,t) = F(q,t)/(1-q)$ leads to the expression
\be
P(q,t) = {1 \over (1-q)} F\left[{e^{-t} \over (1-q)} \right].
\ee
Hence the form of $P$ remains, during the temporal evolution, equal to
the initial condition.
If the initial condition is a $\delta$-function in $q=q_0$, then
$P(q,t) = \delta(q-\langle q \rangle)$ where
$\langle q \rangle = 1-(1-q_0) e^{-t}$.

The fraction of active bonds is then
\be
n_A(t) = 2 \langle q \rangle (1-\langle q \rangle)
= 2 \left[1- (1-q_0) e^{-t} \right] (1-q_0) e^{-t}.
\label{naIsing}
\ee

\begin{figure}
\includegraphics[angle=0,width=10cm,clip]{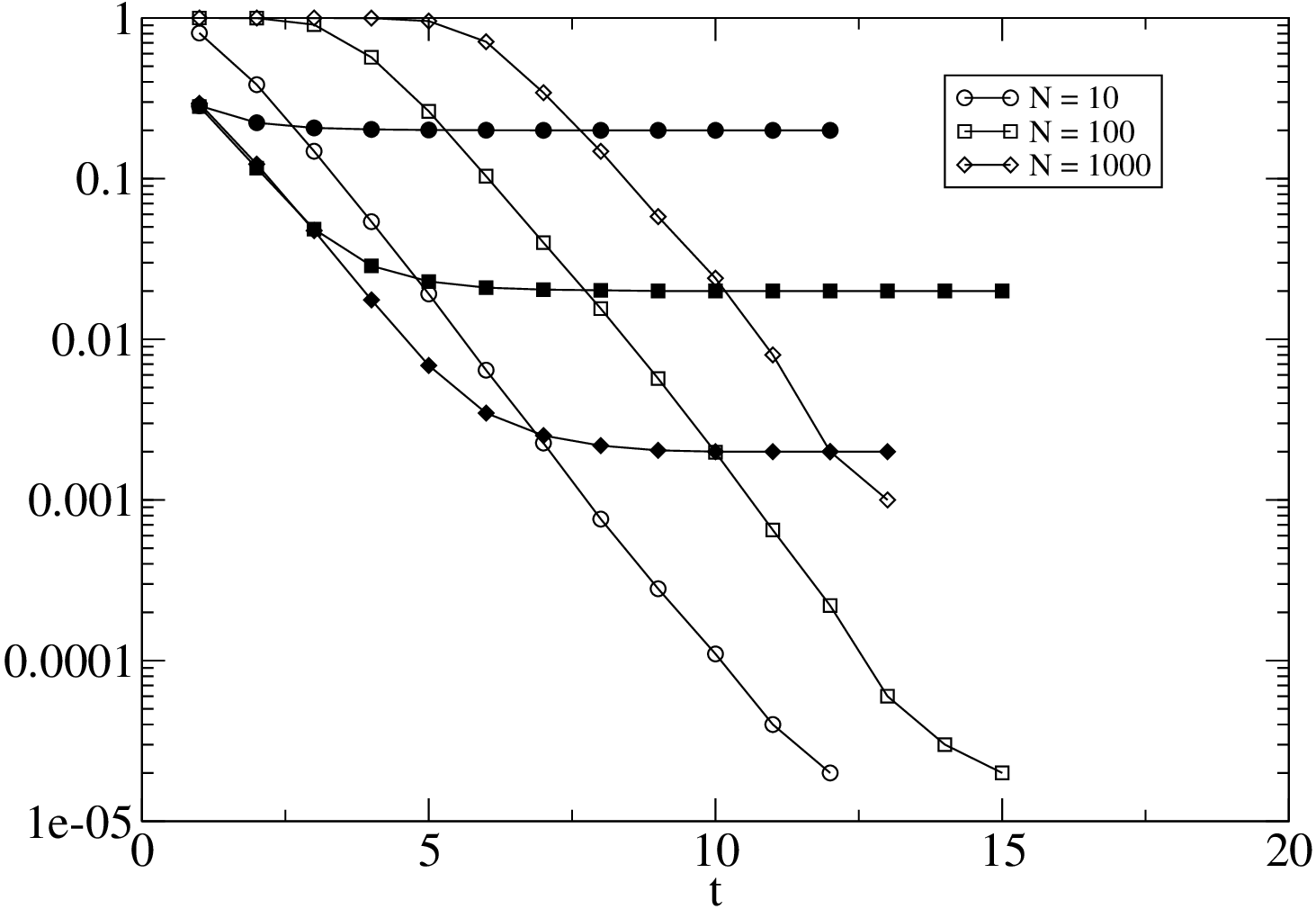}
\caption{Fraction of active bonds in surviving runs $n_A^S(t)$
(filled symbols) and survival probability $\rho(t)$ (empty symbols)
for Glauber $t=0$ dynamics on a complete graph.}
\label{Figx}
\end{figure}

The comparison with numerical simulations
shows perfect agreement already for $N$ of the order of 50.
The exponential decay of $n_A(t)$ is perfectly similar to what happens
for the voter model on a complete graph.
But if we consider separately the fraction of active bonds
in surviving runs $n_A^S(t)$ and the survival probability $\rho(t)$
(Fig.~\ref{Figx}) we find a picture quite different from the case of
voter dynamics [Eqs.~(\ref{eqrho}) and~(\ref{eqnast})].
The fraction of active bonds for surviving run decays exponentially
and then reaches a plateau, but the height of the plateau depends on $N$
and goes to zero as $N \to \infty$.
This is analogous to what occurs on regular lattices and it means that
the Glauber dynamics is effective at ordering the Ising model on a
complete graph.

\subsection{Random graph}

\begin{figure}
\includegraphics[angle=0,width=10cm,clip]{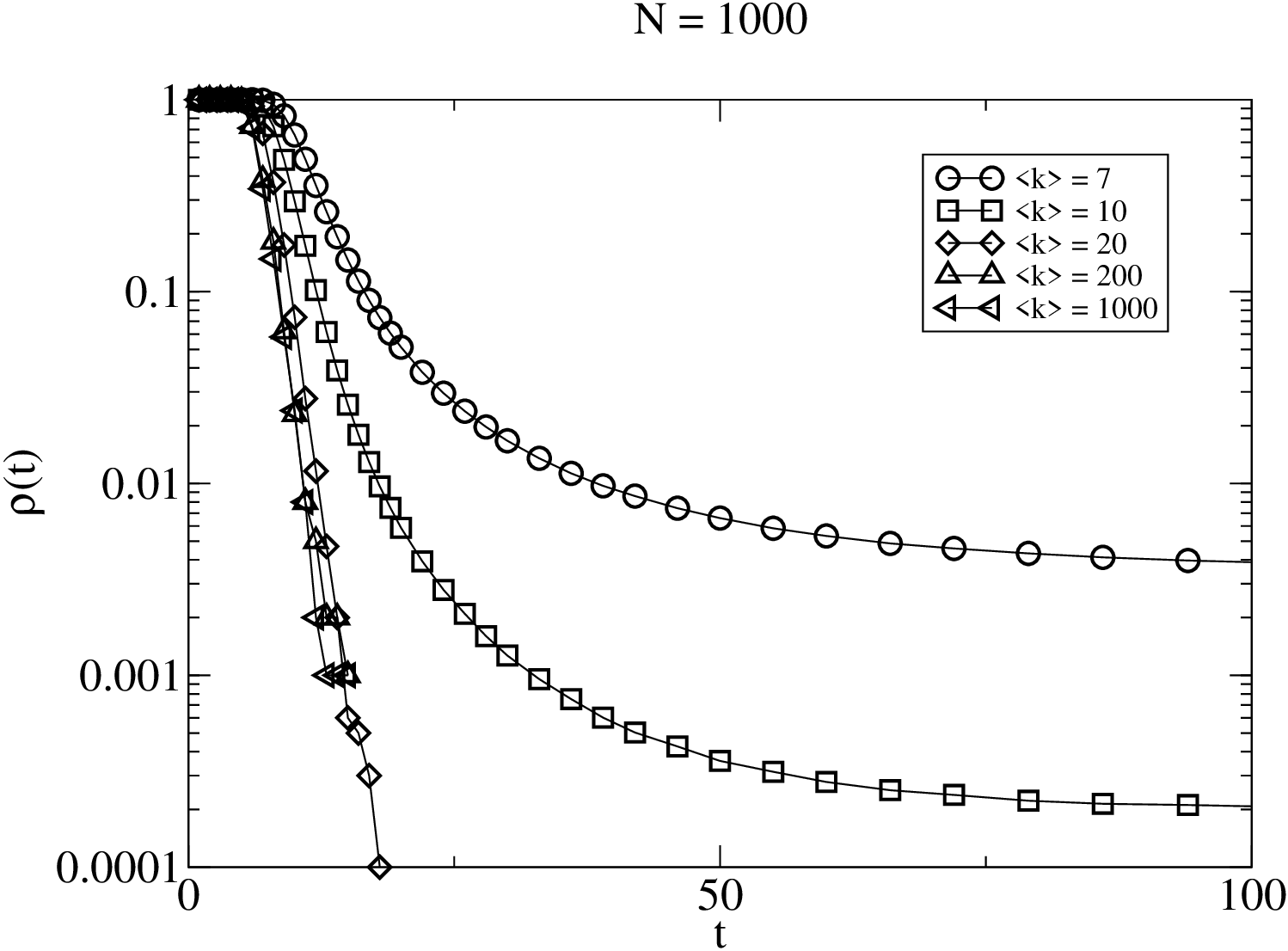}
\caption{Survival probability for Glauber $T=0$ dynamics on a
random graph for different values of the average degree $\km$ of nodes.}
\label{Figx2}
\end{figure}

Let us now consider what occurs for fixed $N$ and changing $p$.
The first change is exhibited by the survival probability
in Fig.~{\ref{Figx2}}.
While for large $\km$ the decay is exponential, for smaller values of
$\km$ a plateau appears, indicating that not all realizations of the
dynamics end up in an ordered state, i.e. with all nodes sharing the
same value of the variable $s_i$.
In such runs the system remains trapped forever in configurations with part
of the nodes with $s_i=-1$ and the rest with $s_i=1$.

A freezing in a disordered state for Glauber dynamics on a random
graph had already been noticed by Svenson~\cite{Svenson01} and considered
analytically by H\"aggstr\"om~\cite{Haeggstroem02}, who showed that,
in the limit $N \to \infty$, the dynamics fails to reach the global
energy minimum (ordered state) for any $\km>0$.
What is the origin of this behavior?

This phenomenon is not related to special realizations
of the random graph topology.
If we fix the topology and let the dynamics evolve many times on it,
we see that in the same finite fraction of runs the system reaches a
disordered state, independently from the particular realization
of the topology.

\begin{figure}
\includegraphics[angle=0,width=10cm,clip]{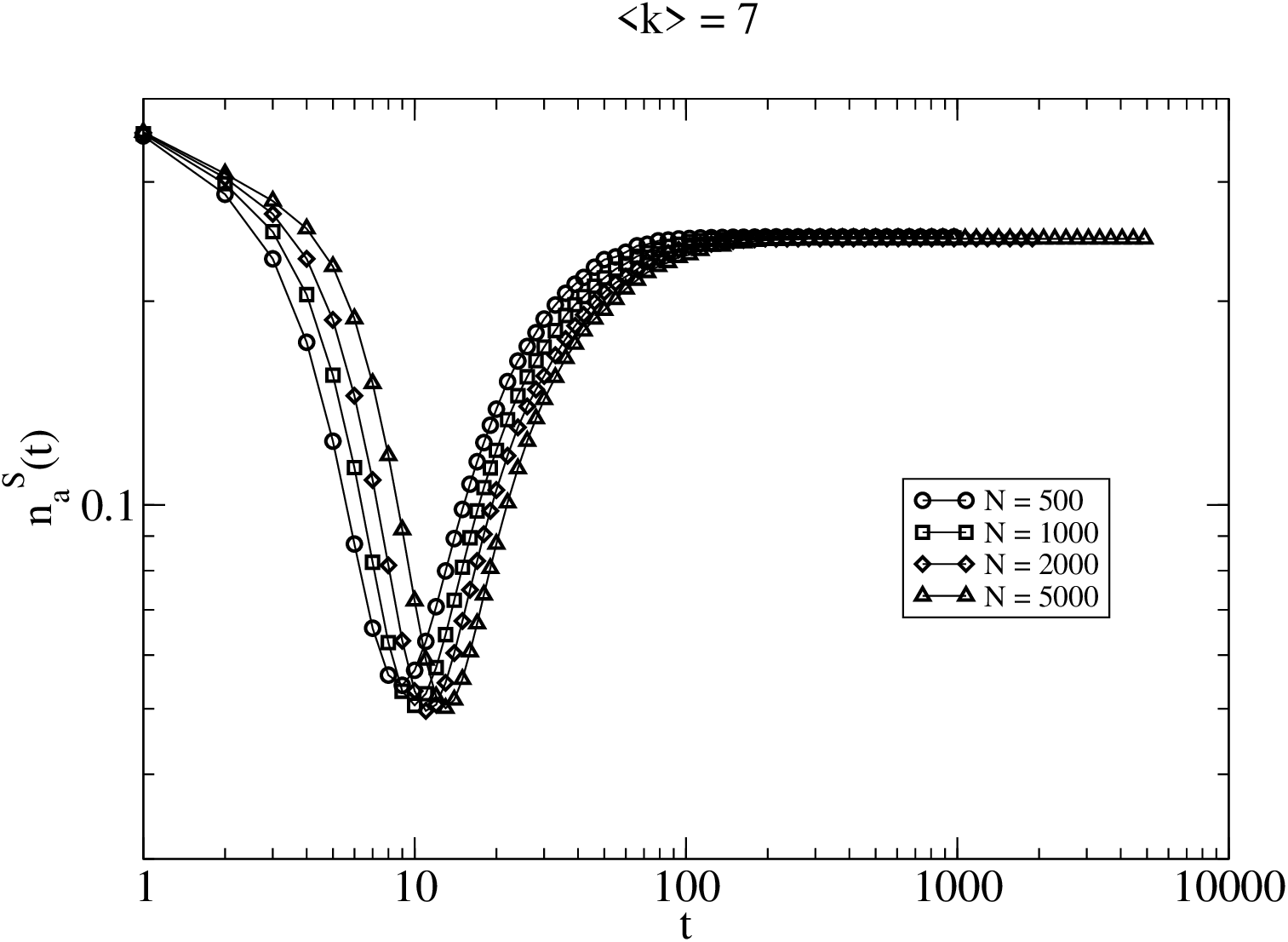}
\caption{Value of the fraction of active bonds in surviving runs $n_A^S(t)$ 
as a function of $t$ for several values of $N$ and $\km=7$.}
\label{Figx3}
\end{figure}

One could think that, given the low value of $p$, there may be small
``communities'' in the graph, i.e. groups of nodes, tightly bound
with each other with only few connections with the rest of the system.
Such communities could become ordered and be basically decoupled from the
rest of the system, leading to a frozen disordered state essentially made
by a huge ordered set and few small chunks ordered in the opposite way.
While this may be true for small $\km$ and large $N$,
here the explanation is different.
The total magnetization ($1/N \sum_i s_i$) in the disordered state
is always very close to zero and the number of domains present is
always equal to 2. 
Moreover the large $t$ limit of the fraction of active bonds in surviving
runs $n_A^S(t)$ indicates that a very high fraction of the bonds
connects sites with different values of $s_i$ (Fig.~\ref{Figx3}).
Hence we can conclude that the system remains trapped
in configurations with two highly intertwined domains of roughly the
same size.
The asymptotic value of $n_A(t)$ is much higher than the lower bound
computed in Ref.~\cite{Haeggstroem02}, $\km e^{-6\km}/256$.

\begin{figure}
\includegraphics[angle=0,width=10cm,clip]{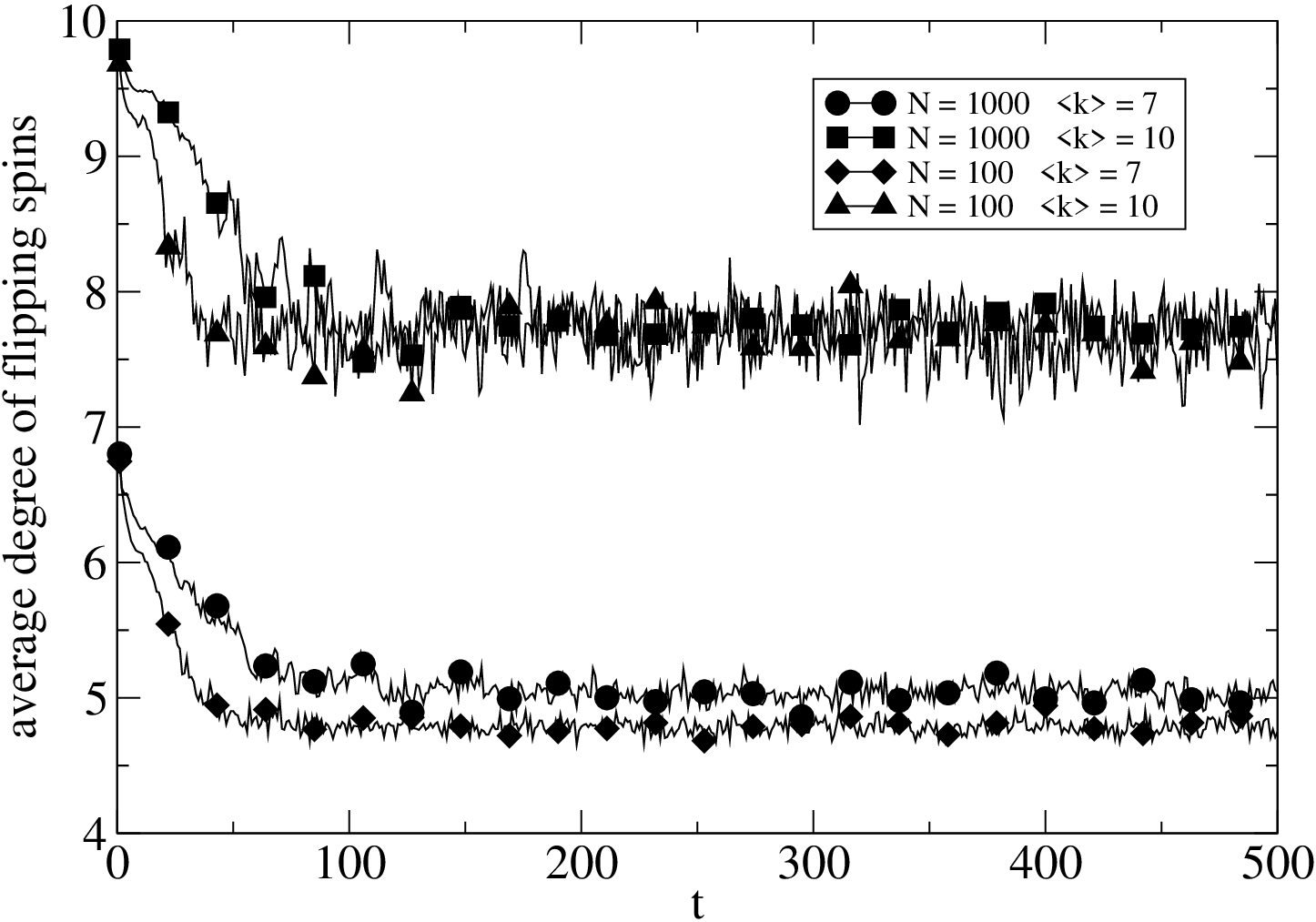}
\caption{Average degree of spins that flip at time $t$ 
for several values of $\km$ and $N$.}
\label{Fig3bis}
\end{figure}

In order to characterize further the dynamics we report in
Fig.~\ref{Fig3bis} the average degree of nodes
that flip at time $t$ for some values of $p$ and $N$.
It turns out that the disordered state is {\em not frozen}.
It is instead a stationary active state, with some spins flipping,
while keeping the energy conserved.
The qualitative picture is then the same holding
on regular lattices for $d>2$: the system wanders forever in an
iso-energy set of states.

\begin{figure}
\includegraphics[angle=0,width=10cm,clip]{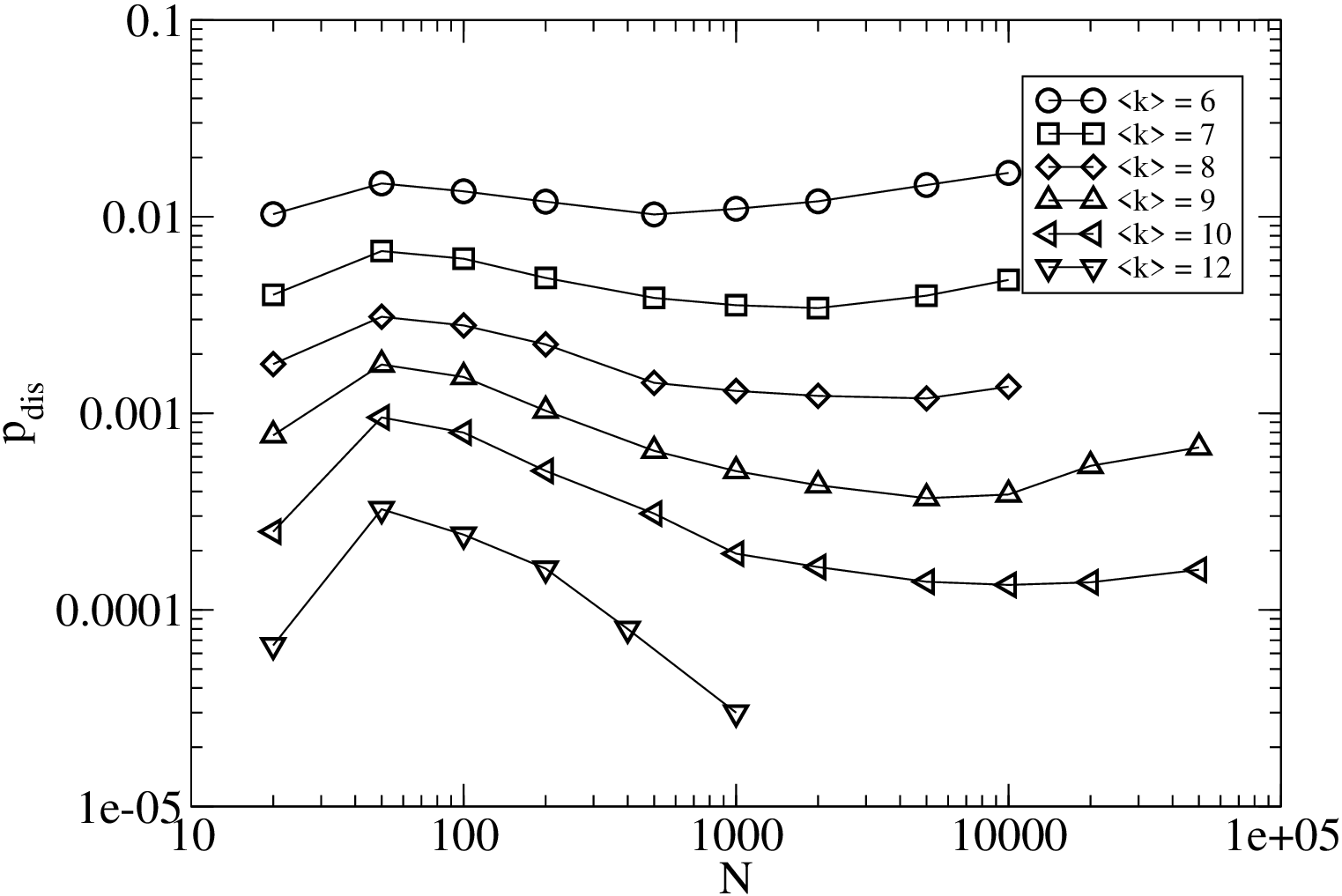}
\caption{Probability of ending in the disordered state
as a function of $N$ for several values of $\km$.}
\label{Fig4}
\end{figure}

It is finally interesting to consider the probability $p_{dis}$
that this stationary disordered state occurs, i.e. the limit
of $\rho(t)$ for $t \to \infty$.
As shown in Fig.~\ref{Fig4}, this probability has a nontrivial behavior
as a function of $N$ for fixed $\km$: for very small and large $N$ it grows,
while it has an intermediate regime such that it decreases as the system
is made bigger. Although Ref.~\cite{Haeggstroem02} guarantees that
$p_{dis}$ goes to 1 as $N$ diverges, for reasonable values of $N$
the values of $p_{dis}$ are much smaller than one.

We have no clear understanding of the reason for the nonmonotonic behavior.
It is probably related in some way to the connectivity transition that
occurs in random graphs for $p=p_1=\ln(N)/N$~\cite{Bollobasbook}.
For $p>p_1$ all nodes belong (in the limit $N \to \infty$) to the giant
component, while for $p<p_1$ separate components exist.
If we invert this relation we obtain an expression for the value of $N$
where the connectivity transition occurs in terms of $\km$:
$N_m = e^{\km}$.
Then for $N<N_m(\km)$ only the giant component exists, while
for $N>N_m(\km)$ some nodes belong to disconnected clusters.
It is tempting to associate $N_m(\km)$ with the value of $N$ such that
$p_{dis}$ is minimal.
The data presented in Fig.~\ref{Fig4} are compatible with this picture,
though we do not have a valid explanation why the decrease of $p_{dis}$
would correspond to the case where only the giant component exists.

\subsection{Barabasi-Albert graph}

On Barabasi-Albert networks, the global behavior of Glauber $T=0$
dynamics  is similar to the one exhibited on random graphs: in a fraction
of the runs the system reaches a disordered stationary state with two domains
of opposite magnetization: spins continue to flip but the energy does not
decrease further.

\begin{figure}
\includegraphics[angle=0,width=10cm,clip]{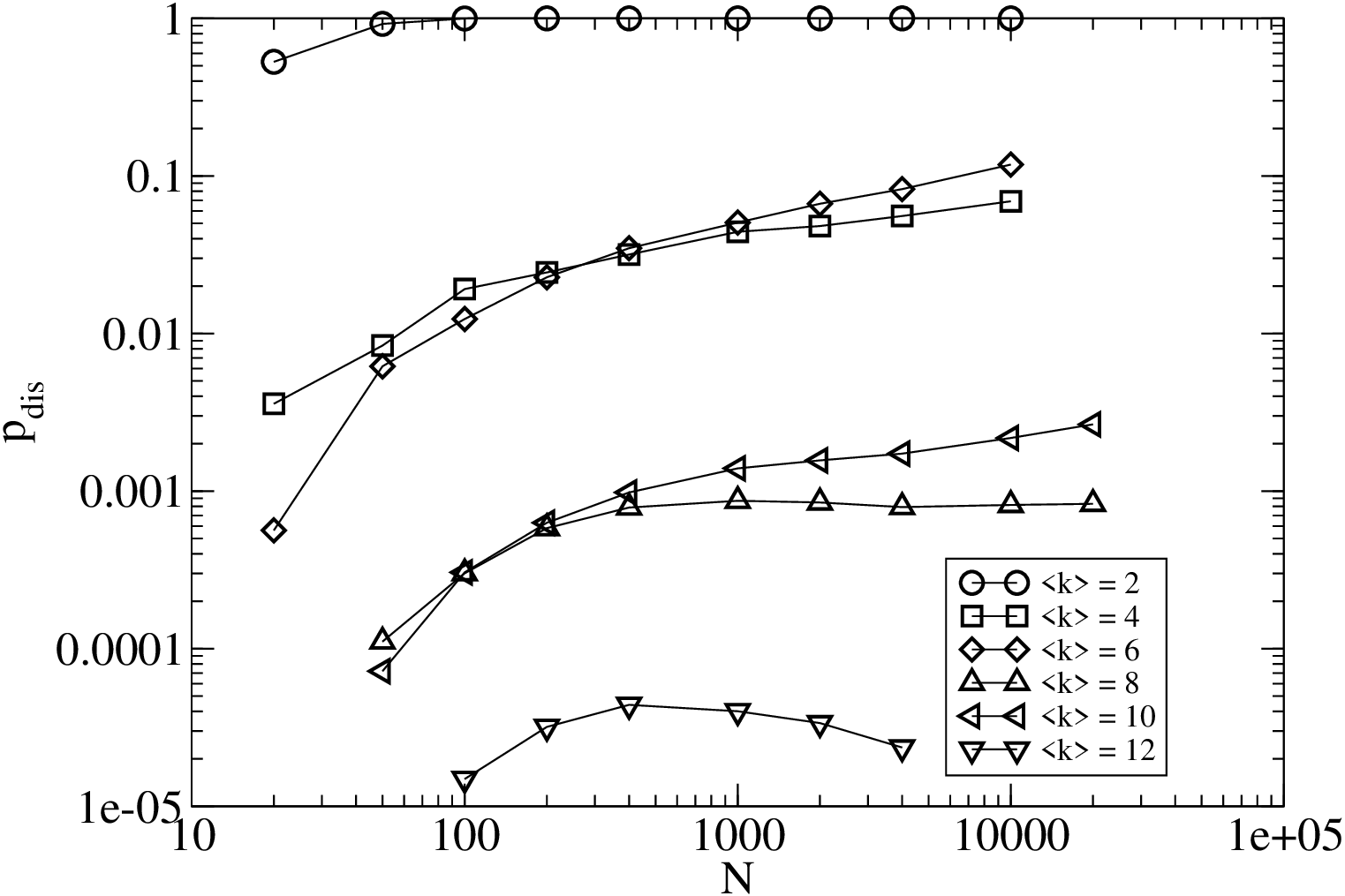}
\caption{Probability that dynamics reaches a disordered state on the
Barabasi-Albert graph as a function of $N$ for the zero-temperature
Glauber dynamics.}
\label{Fig5}
\end{figure}

In Fig.~\ref{Fig5} we report the fraction of such runs as a function
of the number of nodes $N$, for several values of the average degree $\km=2m$,
where $m$ is the number of edges added for each new node.
In analogy with the case of the random graph, the probability of remaining
disordered grows for large values of $N$ except for the case $\km=12$, where
a decrease is seen. This is similar to what happens on a random graph for
large average degree.
For $\km=2$ the probability rapidly reaches the value 1, i.e. no run 
reaches full order.
This can be easily understood given the tree-like structure of the
Barabasi-Albert graph for $m=1$.

\section{Conclusions}

In summary, we have investigated the behavior of the
simplest ordering dynamics for a two-valued variable on networks
ranging from the fully connected graph to random and scale-free graphs.
In general we find that the difference between the zero-temperature
Glauber-Metropolis Ising dynamics and the voter model has a quite strong
impact on the ordering process starting from a completely random
initial condition.
On the other hand, the presence or absence of a typical scale in the 
network describing the interaction pattern has a limited effect:
it changes (in the voter model) the temporal scale over which order
is reached, but does not affect whether or not such order is reached.

The voter model dynamics invariably leads to full ordering, for any
type of topology, provided the number $N$ of nodes in the system is finite.
When $N$ grows the time needed to reach complete
consensus diverges, in a way that depends on the connectivity pattern.
If one considers the number of active bonds $n_A(t)$ as a function of time
averaged over all realizations of the dynamics, including those that
already have reached the fully ordered state, one sees an exponential
decay. This may lead to the conclusion that the the system actually orders
exponentially fast (i.e. faster than on regular lattices).
However, this conclusion disagrees with the connection between
the voter dynamics and the recurrence properties of random walks.
The recurrence of the random walk on regular lattices for $d \le 2$
implies that the voter model orders on them, while it remains in
a disordered state when the walk is transient ($d>2$).
The same argument implies that the voter model does not order on
networks, as those considered here, for which the random walk is
transient~\cite{Bollt04}.
The solution of this apparent paradox is that the voter model
actually does not get ordered on networks in the thermodynamical limit.
The right quantity to observe this is the density of active bonds in
surviving runs $n_A^S(t)$, which does not decay to zero; it
attains large values, signaling that in surviving runs the
system is again split in two domains with a large number of
interconnections.
Notice that this is true also for the complete graph.

For Glauber dynamics on a complete graph full ordering is attained
for any system size, including the thermodynamic limit.
Randomness in the connectivity pattern implies instead that
even a finite system has a nonzero probability to remain trapped in metastable
states, i.e. to indefinitely cycle through configurations with the same energy.
In such metastable configurations, the system is split in two domains
with a large number of interconnections.
When the system size grows the probability of reaching this disordered
stationary state tends to increase, making full ordering less likely.
Apart from details, this holds true for both random (Erd\"os-Renyi)
and scale-free (Barabasi-Albert) graphs.

We are now in the position to summarize the common features and the
differences between the behavior of the voter model and the Glauber
zero-temperature dynamics.
At the level of the complete graph, despite the apparent similarity
between Eq.~(\ref{navoter}) and Eq.~(\ref{naIsing}), the two models
are different, since Glauber reaches genuine order in the thermodinamic
limit, while voter does not.
In the presence of a random topology, the similarity is stronger:
In both cases, the system initially approaches a state with
two intertwined domains of roughly the same size. For the voter model,
the large noise present in the dynamics creates fluctuations
that eventually lead, in finite systems, to complete ordering.
In the Glauber case, instead, the zero-temperature condition forbids
energy fluctuations and the dynamics remains confined to isoenergetic
partially ordered configurations.

The two dynamical models considered are not dramatically sensitive to the
underlying topology. The existence or absence of a characteristic scale
in the degree distribution of the network does not affect whether
order is reached or not.
For the voter dynamics, the degree distribution
only affects the way the characteristic temporal scale $\tau$ depends on $N$.
We have shown that for over three decades the exponent $\gamma = 0.880 \pm 003$
fits very well the numerical data, but this does not rule out the analytical
prediction $\tau \sim N/\log N$ of Ref.~\cite{Sood04}.
However, the results for the scale-free graph by Dorogovtsev, Mendes and
Samukhin seem to indicate that the precise value of $\gamma$ may
depend on other details of the underlying network.

\section{Acknowledgments}

This research has been supported by the ECAGENTS project founded by the
Future and Emerging Technologies program (IST-FET) of the European
Community under EU R\&D contract IST-2003-1940, and by the project DELIS
(contract 001907).
The information provided
is the sole responsibility of the authors and does not reflect the
Community's opinion. The Community is not responsible for any use that
may be made of data appearing in this publication.

\end{document}